\documentclass[aps,prl,twocolumn,superscriptaddress,showpacs,floatfix]{revtex4}
\usepackage{graphicx,epsfig,amsmath}

\newcommand{\subfig}[2]{Fig.~\ref{fig:#1}(#2)}

\begin{document}


\title{Measuring Charge Transport in an Amorphous Semiconductor Using Charge Sensing}

\author{K. MacLean}
    \email{kmaclean@mit.edu}
    \affiliation{Department of Physics, Massachusetts Institute of Technology, Cambridge, Massachusetts 02139}
\author{T. S. Mentzel}
    \affiliation{Department of Physics, Massachusetts Institute of Technology, Cambridge, Massachusetts 02139}
\author{M. A. Kastner}
    \affiliation{Department of Physics, Massachusetts Institute of Technology, Cambridge, Massachusetts 02139}


\begin{abstract}
We measure charge transport in hydrogenated amorphous silicon (a-Si:H) using a nanometer scale silicon MOSFET as a charge sensor. This charge detection technique makes possible the measurement of extremely large resistances. At high temperatures, where the a-Si:H resistance is not too large, the charge detection measurement agrees with a direct measurement of current.  The device geometry allows us to probe both the field effect and dispersive transport
in the a-Si:H using charge sensing and to extract the density of states near the Fermi energy.
\end{abstract}


\maketitle

A variety of technologically promising materials and devices, from arrays of semiconducting nanocrystals, that are candidates for solar energy harvesting \cite{klimov:mult}, to lateral quantum dots that hold potential for quantum information processing \cite{Petta2005:CoherentManipulation}, are highly resistive \cite{Us:CdSe,Mentzel:PbSe}.   For the latter, charge measurement using a sensor integrated with the device \cite{Field1993:NoninvasiveProbe} has recently been widely utilized to probe quantum mechanical phenomena that would be impossible to observe by measuring current \cite{ChargeSensingGaAs}.  For the study of highly resistive materials scanning probe techniques have been used to determine the charge distribution \cite{YacobyQImage,Drndic:EFM}. However, a study of the charge transport properties of a resistive material using an integrated charge sensor has yet to be realized.

In this Letter, we illustrate the power of this charge sensing technique by investigating transport in hydrogenated amorphous silicon (a-Si:H). By patterning a strip of a-Si:H thin film adjacent to a nanometer scale silicon MOSFET, we are able to detect charging of the a-Si:H and measure extremely high resistances ($\sim 10^{17}$ $\Omega$) using moderate voltages ($\sim$ 1 V). We compare our results with those of current measurements at high temperatures, where the resistance is not too large, and find good agreement. The two methods complement each other in that they probe different ranges of electrical resistance. Our device geometry, in which the MOSFET sensor and a-Si:H can be gated independently, allows us to investigate a variety of transport phenomena, including the field effect \cite{Cohen:DOS} and dispersive transport \cite{Kastner:Dispersive,Rose:Dispersive}, using charge sensing \cite{Senturia:CFT,CFTfootnote}.  We use these methods to probe the density of localized states near the Fermi energy, and obtain consistent results.  Our method constitutes a new resistance measurement, which is insensitive to the resistance of the contacts supplying the charge, and should be applicable to a wide variety of materials. 

\begin{figure}

\begin{center}
\includegraphics[width=9.0cm, keepaspectratio=true]{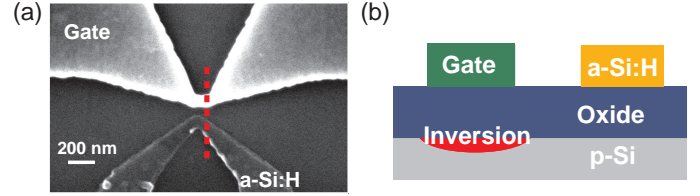}
\end{center}

\caption{(a) Electron micrograph of MOSFET gate and a-Si:H strip.  A positive voltage is applied to the MOSFET gate, forming an inversion layer underneath. (b) Vertical sketch of the device geometry along the dashed line in (a).  The conductance through the inversion layer formed under the gate is sensitive to the charge on the a-Si:H strip.}
\label{fig:fig1}
\end{figure}

Our charge sensor consists of an $n$-channel MOSFET that is electrostatically coupled to a strip of a-Si:H.  An electron micrograph of the structure is shown in \subfig{fig1}{a}.  The MOSFET is fabricated using standard techniques \cite{CMOS} on a $p$-type Silicon substrate. 
The $n^{+}$ polysilicon gate of the MOSFET is patterned using electron beam lithography and reactive ion etching and tapers down to a width of $\approx$ 60 nm.  The characteristics of this narrow channel MOSFET are similar to those reported previously \cite{Kastner:NarrowMos}.  Because of its narrow width, the MOSFET is extremely sensitive to its electrostatic environment \cite{Ralls:Switch}. Furthermore, it has a relatively thick gate oxide (thickness $d_{ox}$ = 100 nm), which ensures that the metallic polysilicon gate does not effectively screen the inversion layer from nearby electrostatic fluctuations.  Adjacent to the MOSFET, we pattern a strip of phosphorous doped a-Si:H.  The narrowest portion of the strip is located $\approx$ 70 nm from the narrowest portion of the MOSFET. 

The a-Si:H is deposited by plasma enhanced chemical vapor deposition \cite{aSiH}, with a gas phase doping ratio and hydrogen dilution of [PH$_{3}$] / [SiH$_{4}$] = $2 \times 10^{-2}$ and [H$_{2}$] / [SiH$_{4}$ + H$_{2}$] = 0.5, respectively.  Because we use a relatively large doping level, we expect a large defect density 
$N_{D} \sim$ 10$^{18}$ cm$^{-3}$ \cite{aSiH}. The deposition substrate temperature is $T_{s}$ = 200 C, and the deposition rate is $\approx$ 0.17 nm/s.  The a-Si:H film is patterned using electron beam lithography and a lift-off technique, which will be described in a subsequent publication \cite{MacLean:FutureCh}.  The a-Si:H strip is connected to two gold contacts (separated by $\approx$ 2 $\mu$m), and the MOSFET inversion layer is contacted through two degenerately doped $n^{+}$ regions, none of which are shown in \subfig{fig1}{a}. After sample preparation, the device is loaded into a cryostat, and kept in Helium exchange gas throughout the course of the measurements discussed here.

The parameters used to deposit our a-Si:H are similar to those studied previously,
and thicker a-Si:H films than those used for our charge sensing measurements, deposited under the same conditions, have conductivities and activation energies similar to those reported elsewhere \cite{aSiH}. However, the sample studied in this work is nanopatterned ($\approx$ 100 nm wide at its narrowest point) and also is only $\approx$ 50 nm thick.   Thus, although the characteristics reported below are similar to what one would expect for thick heavily doped films, we expect that for our sample surface effects may be significant \cite{Fritzsche:Surface}, and there may also be differences in morphology and hydrogen content as compared with thicker films.

\begin{figure}

\begin{center}  
\includegraphics[width=6.5cm, keepaspectratio=true]{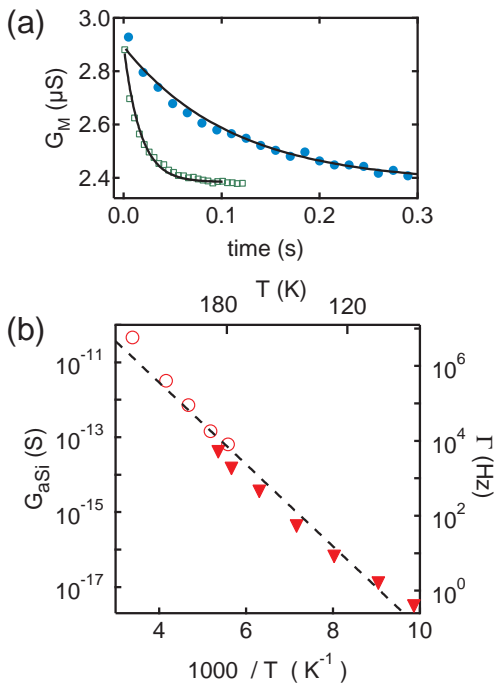}
\end{center}
    \caption{(a) $G_{M}$ as a function of time at T = 125 K (closed circles) and 140 K (open squares), the latter is offset for clarity. At $t = 0$ the voltage applied to one of the a-Si:H contacts is changed from -1.8 V to -2.7 V. For these traces multiple charge transients have been averaged to improve signal-to-noise.  The solid lines are theoretical fits described in the main text. (b) Conductance $G_{aSi}$ obtained from charge transients (closed triangles) and direct conductance measurements (open circles), measured with $V_{ds}\approx$ -2.3 V. For the charge transient measurement, $\Gamma$ is given on the right hand axis. The dashed line is a theoretical fit described in the main text.}
    \label{fig:fig2}
\end{figure}

	Our measurement consists of monitoring the MOSFET conductance $G_{M}$ as a function of time after changing the voltage applied to one of the a-Si:H contacts (\subfig{fig2}{a}).  We set the voltage of one of the a-Si:H contacts to 0 V (relative to the $p$-type substrate), and, at $t=0$, rapidly change the voltage applied to the other a-Si:H contact from -1.8 V to -2.7 V.  This causes additional electrons to move onto the a-Si:H strip from the gold contacts.
The MOSFET senses this change in charge electrostatically, and $G_{M}$ decreases.  This effect is shown in \subfig{fig2}{a}.  Following the voltage step, $G_{M}$ decreases with time
at a rate that increases when the temperature is increased. A similar decrease in $G_{M}$ is observed when the step is applied to the other a-Si:H contact or to both contacts simultaneously. 
This decrease in $G_{M}$ is caused by the charging of the a-Si:H strip capacitor, and we henceforth refer to it as a charge transient \cite{sepdev}. When the a-Si:H contact voltage is changed back to its original value, the same transient is observed but with the opposite sign.

	For a resistive film with a distributed capacitance, the charge stored as a function of position and time $\sigma(x,t)$ obeys the diffusion equation, with a diffusion constant given by $D^{-1} = C R_{sq} $ \cite{AgarwalTSCD,Drndic:EFM}.  Here $C$ is the capacitance per unit area between the a-Si:H and the underlying substrate, and $R_{sq}$ is the resistance per square of the a-Si:H film. Because $C$ is reduced by any depletion in the $p$-type silicon underneath the a-Si:H, $C$ will vary with the voltages applied to the MOSFET gate, a-Si:H contacts, and $p$-type substrate.  However, we estimate that for the range of voltages used to collect the data reported below, $C$ remains within a factor of 5 of the oxide capacitance ($C_{ox} = \kappa_{ox} \epsilon_{0}/ d_{ox}$), and, in view of the large variations in $D$ reported below, can be treated as a constant \cite{Capacitance}.  
	
	For a strip of material of length $L$, for which the potential at one end is changed from $V$ to $V + \Delta V$ at $t$ = 0, the charge at any point along the strip varies exponentially with time as $\sigma(t) \approx \sigma_{\infty} - \sigma_{\Delta} e^{-\Gamma t}$ for sufficiently large $t$. Here $\Gamma = \pi^{2} D / L^{2}$, and $\sigma_{\infty}$ and $\sigma_{\Delta}$ are constants that depend on $V$, $\Delta V$, and C. 
For a sufficiently small voltage step $G_{M}$  varies linearly with $\sigma$, so
$G_{M}(t) \approx G_{\infty} + G_{\Delta} e^{-\Gamma t}$, where the sign of $G_{\Delta}$ is opposite to the sign of $\Delta V$.  The solid curves shown in \subfig{fig2}{a} are fits to this equation, from which we extract $\Gamma$ \cite{Diffusion2}.

From our measurement of $\Gamma$ and the values of $L$ and $C$ for our a-Si:H strip, we extract $R_{sq}$, and from this compute the conductance $G_{aSi} = w/(R_{sq} L)$, where $w$ is width of the a-Si:H strip.  In \subfig{fig2}{b} we plot $G_{aSi}$ and $\Gamma$ and as functions of temperature. $G_{M}$ is weakly temperature dependent, and for this data we therefore adjust the MOSFET gate voltage as we vary the temperature to keep $G_{M}$ approximately constant. At higher temperatures we are able to directly measure $G_{aSi} = dI/dV_{ds}$, where $V_{ds}$ is the voltage between the a-Si:H contacts, and these results are also shown in \subfig{fig2}{b}.  At T $\approx$ 180 K, we can measure $G_{aSi}$ using both techniques, and the results are in good agreement.  The measurements are complementary, in that the charge transient technique is easier to implement for smaller conductances $G_{aSi}$ because the charging is slower, while a measurement of current is only possible for larger values $G_{aSi}$. The dashed line in \subfig{fig2}{b} is a fit to $G_{aSi}(T) = G_{0}e^{-E_{a}/kT}$.  The data are thus consistent with an activated transport mechanism, with an activation energy $E_{a} \approx$ 200 meV, as is typically observed for a-Si:H films heavily doped with phosphorous \cite{aSiH}. We note that at the lowest temperatures, we measure resistances as high as $\sim$ 10$^{17}$ $\Omega$.

\begin{figure}
\setlength{\unitlength}{1cm}
\begin{center}
\includegraphics[width=6.5cm, keepaspectratio=true]{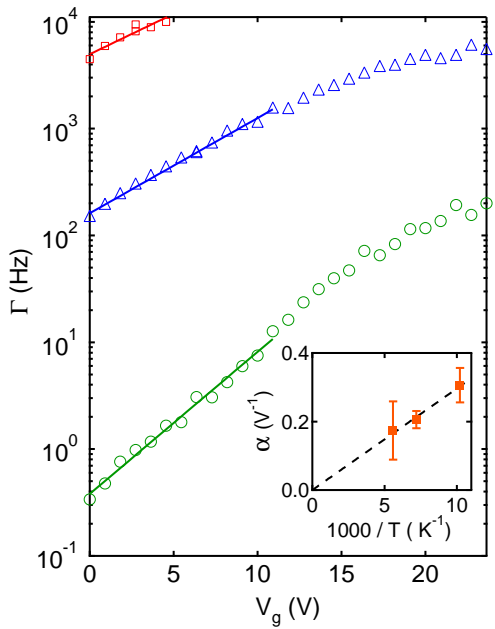}
\end{center}
\caption{$\Gamma$ as a function of $V_{g}$ at T = 98 K (circles), 139 K (triangles), and 179 K (squares). The solid lines are theoretical fits described in the main text. (Inset) $\alpha$ as a function of inverse temperature.  The dashed line is a theoretical fit described in the main text.}
    \label{fig:fig3}
\end{figure}

We can also measure $\Gamma$ as a function of gate voltage (Fig. \ref{fig:fig3}). For this measurement, we apply the same voltage $V_{aSi}$ to both a-Si:H contacts relative to the $p$-type substrate.  The effective gate voltage is then $V_{g} = - V_{aSi}$.  We then add a small voltage step $\Delta V \approx 0.5 V$ to $V_{aSi}$ to produce a charge transient, from which $\Gamma$ is extracted as in \subfig{fig2}{a}.  For all of the measurements shown here the substrate voltage is held constant at -3 V, and voltages are quoted relative to this value. Unlike previous reports \cite{Senturia:CFT}, our geometry allows us to maintain an approximately constant value for the MOSFET conductance, and thus to maintain a high charge sensitivity, as we make large changes in $V_{aSi}$ by applying smaller compensating voltage shifts to the MOSFET gate voltage. This allows us to perform this field effect measurement for a large range of sample conductances.

In Fig. \ref{fig:fig3} we plot $\Gamma$ as a function of $V_{g}$ at three different temperatures. We see that $\Gamma$ rises with $V_{g}$, indicating $n$-type conduction through the a-Si:H, as expected for a phosphorous doped sample. The exponential increase in $\Gamma$ with $V_{g}$ is consistent with the activated
conduction found in \subfig{fig2}{b}, provided we assume an approximately constant density of localized states. We have:
\begin{equation}\label{eq:exp}
		\Gamma = \omega_{0}e^{-E_{A}/kT}
\end{equation}
Here $\omega_{0}$ is a prefactor that depends only weakly on temperature, and the activation energy $E_{A}$ is reduced as the gate voltage moves the Fermi level closer to the mobility edge.  The logarithmic slope $\alpha = \partial ln(\Gamma)/\partial V_{g}$ is then given by $\alpha = \frac{1}{kT}\partial E_{A}/\partial V_{g} = C / (ekT\rho(E_{F})s_{tf})$, where $\rho(E_{F})$ and $s_{tf}$ are the density of states at the Fermi energy and Thomas-Fermi Screening length, respectively \cite{Mentzel:PbSe,zeroTGate}.  Thus we expect an exponential increase in $\Gamma$ with $V_{g}$ as long as the product $\rho(E_{F})s_{tf}$ is constant.

At each temperature, we fit the data to obtain $\alpha$ (solid lines in Fig. \ref{fig:fig3}), and, in the inset
to Fig. \ref{fig:fig3}, we plot $\alpha$ as a function of inverse temperature. The dashed line is a linear fit (constrained to pass through zero) and is consistent with the data. From the slope of this fit we obtain 
$s_{tf}\rho(E_{F})\approx$ 5 $\times$ 10$^{13}$ eV$^{-1}$cm$^{-2}$, and, expressing $s_{tf}$
in terms of $\rho(E_{F})$ and the a-Si:H dielectric constant \cite{Mentzel:PbSe}, we solve for $\rho(E_{F})\sim$ 10$^{20}$ eV$^{-1}$cm$^{-3}$.  The density of states at the Fermi level
for phosphorous doped amorphous hydrogenated silicon obtained from more commonly used transport techniques
is typically $\sim$ 10$^{19}$ eV$^{-1}$cm$^{-3}$ \cite{Cohen:DOS}.  The fact that our $\rho(E_{F})$ is somewhat high is not surprising, given the large gas phase doping level used in our a-Si:H film deposition. For the range of voltages used here $E_{F}$ moves by an amount comparable to values of the band tail width commonly found for a-Si:H films \cite{aSiH}, so we expect that $\rho(E_{F})$ should increase somewhat as $V_{g}$ is made more positive and the Fermi level is moved in the band tail. This may cause the observed decrease in logarithmic slope $\alpha$ for $V_{g} > $15 V  in Fig. \ref{fig:fig3}.

\begin{figure}[!]
    \begin{center}
        \includegraphics[width=8.0cm, keepaspectratio=true]{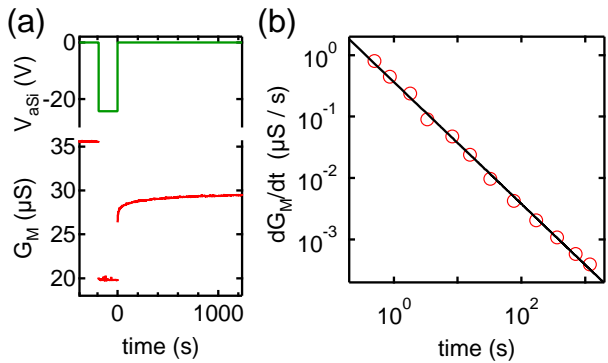}
    \end{center}
    \caption{(a) Voltage applied (top panel) and transistor conductance (lower panel) used to measure dispersive transport at T = 89 K, as discussed in the main text (b) $\partial G_{M}(t)/\partial t$ extracted from the data shown in (a).  The solid line is a theoretical fit described in the main text.}
    \label{fig:fig4}
\end{figure}

	At lower temperatures, where the time scale for charging is longer, we observe dispersive transport 
\cite{Kastner:Dispersive,Rose:Dispersive} (\subfig{fig4}{a}).  When we step $V_{aSi}$ from 0 V to -24 V, $G_{M}$ quickly drops.  However, when $V_{aSi}$ is stepped back to 0 V, $G_{M}$ rises at slower rate, and does not regain its original value.  This behavior can be understood as follows:  After the negative $V_{aSi}$ step the a-Si:H quickly charges, as electrons can enter the a-Si:H at energies close to the conduction band.  However, as time progresses, these electrons get trapped in localized states deeper in the band gap. When the voltage is returned to its original value, the a-Si:H therefore takes a much longer time to discharge: From Eq. \eqref{eq:exp}, the time necessary to release electrons from states at an energy $E_{A}$ below the transport energy is $t\sim \Gamma^{-1} = \omega_{0}^{-1}e^{-E_{A}/kT}$.  As electrons deeper and deeper in the gap are released, $t$ grows, and thus the transport process becomes dispersive \cite{Kastner:Dispersive,Rose:Dispersive}.

This can be made quantitative: At a time $t$ after the negative voltage step, only electrons in localized states with energies $E_{A} < E_{max} = kTln(\omega_{0}t)$ \cite{Kastner:Dispersive,Rose:Dispersive} are able to escape from the a-Si:H. The charge on the a-Si:H is then given by $\sigma (t) = e\int^{E_{max}} s_{tf}\rho(E_{A})dE_{A}$ (up to an additive constant).  Assuming a constant density of states and differentiating with respect to time we obtain:
	
\begin{equation}\label{eq:Disp}
		\partial \sigma /\partial t =  e s_{tf}\rho(E_{F})kT/t
\end{equation}

In \subfig{fig4}{b} we plot the derivative of $G_{M}$ with respect to time on a log-log plot: A fit to a power law dependence (solid line) yields a power of -1 $\pm$ 0.1 as expected. Moreover, the prefactor of this power law is $e s_{tf}\rho(E_{F}) r$, where $r$ is the conversion between $\sigma$ and $G_{M}$ that can be estimated from the  decrease in $G_{M}$ after pulsing $V_{aSi}$ to -24 volts.  We obtain $\rho(E_{F}) \sim 10^{20}$ eV$^{-1}$cm$^{-3}$, consistent with the value extracted from the data in Fig. \ref{fig:fig3}.
	
	There are some aspects of our charge transient technique that are not fully understood.  While the measurements of $\Gamma$ shown in Fig. \ref{fig:fig2} and Fig. \ref{fig:fig3} do not depend strongly on the voltage applied between the a-Si:H contacts $V_{ds}$ for $V_{ds} <$ 1, we observe a large nonlinearity in $G_{aSi}$ at room temperature when we measure current as a function of voltage at $V_{ds} \approx$ 500 mV. While the source of this disagreement is unclear, it is possible that at zero bias the steady state current is limited by the narrow constriction or by the contacts. The charge detection method is not sensitive to such effects because it only requires that charge move in the a-Si:H, and not that the charge flow continuously through the entire sample. Indeed, our charge sensing method is effective even in the presence of blocking contacts; it can detect charge diffusing toward the contact even with infinite contact resistance as long as there is significant contact capacitance. Future work will study in detail the sensitivity of our technique to contact resistance. 
	
		In summary, integration of a charge sensor provides a new technique that makes it possible to characterize the charge transport properties of extremely resistive materials, even with poor contacts. While we have chosen a-Si:H as an example, we expect that the technique will be useful for a wide variety of materials and systems. 
	
	We are grateful to L. Levitov, H. Fritzsche, J. Kakalios, Iuliana P. Radu, and S. Amasha for discussions.  This work was supported by the US Army Research Office under Contract W911NF-07-D-0004 and the Department of Energy under Award Number DE-FG02-08ER46515.


\begin{thebibliography}{25}
\bibitem[{CFT()}]{CFTfootnote}
\bibinfo{note}{Our device is somewhat similar to the charge flow transistor
  \cite{Senturia:CFT} pioneered by Senturia and coworkers. However, our device
  geometry enables the study of a much wider range of transport phenomena, as
  is discussed in the main text.}
\bibitem[{sep()}]{sepdev}
\bibinfo{note}{For a separate device, where only one of the two gold contacts
  is connected to a-Si:H close to the MOSFET, we observe a charge transient only
  when the pulse is applied to the contact connected to the a-Si:H adjacent to the
  MOSFET, verifying that the charge transient is indeed caused by changes in
  the a-Si:H charge.}
\bibitem[{Cap()}]{Capacitance}
\bibinfo{note}{For all calculations depending on $C$, we use the midpoint of
  this range, $C = 3C_{ox}/5$}.
\bibitem[{Dif()}]{Diffusion2}
\bibinfo{note}{We have checked that leaving out the faster terms in the full
  solution to the diffusion equation does not significantly effect our results
  \cite{MacLean:FutureCh}. Because the results from this charge diffusion model
  do not depend on the width of the strip, including the gradual taper of the
  width of the strip of a-Si:H studied in our experiment in our theoretical model
  is also unnecessary.}
\bibitem[{zer()}]{zeroTGate}
\bibinfo{note}{For these expressions we have used the T = 0 limit, which is
  justified provided $\rho(E)$ does not vary substantially over a range of
  energy kT about $E_{F}$ \cite{Cohen:DOS}.}
\end{thebibliography}

\end{document}